# Requirements for Automated Assessment of Spreadsheet Maintainability

José Pedro Correia, Miguel A. Ferreira
Software Improvement Group,
Rembrandt Tower, 14th floor
Amstelplein 1, 1096 HA Amsterdam, The Netherlands
Email: j.p.correia@sig.eu, m.ferreira@sig.eu

## ABSTRACT

*The use of spreadsheets is widespread. Be it in business, finance, engineering or other areas, spreadsheets are created for their flexibility and ease to quickly model a problem. Very often they evolve from simple prototypes to implementations of crucial business logic.*

*Spreadsheets that play a crucial role in an organization will naturally have a long lifespan and will be maintained and evolved by several people. Therefore, it is important not only to look at their* reliability, *i.e., how well is the intended functionality implemented, but also at their* maintainability, *i.e., how easy it is to diagnose a spreadsheet for deficiencies and modify it without degrading its quality.*

*In this position paper we argue for the need to create a model to estimate the maintainability of a spreadsheet based on (automated) measurement. We propose to do so by applying a structured methodology that has already shown its value in the estimation of maintainability of software products. We also argue for the creation of a curated, community-contributed repository of spreadsheets.*

## 1 INTRODUCTION

Although there is a considerable amount of work in the literature regarding the occurrence of errors in spreadsheets, another important aspect has been so far mostly neglected, namely *maintainability*. Most surveys conducted to determine the degree of usage of spreadsheets in large companies agree on one thing: their presence is almost ubiquitous. Furthermore, they very often play crucial roles in the organizations where they are used [Croll, 2009].

Much like software systems, their life span can be of many years and they can be developed and maintained by several different people. This makes it crucial to view spreadsheets from a different perspective, namely to study how easy they are to maintain, *i.e.*, how easy it is for the person responsible for a spreadsheet to:

- understand it, diagnose it for deficiencies or determine where change is required;
- actually perform a required change;
- avoid unintended effects of a local change in the rest of the spreadsheet;
- validate the modified spreadsheet.

There is latent risk in a spreadsheet that is difficult to maintain, since errors will be more difficult to identify and/or address, and errors are extremely common to occur [Panko, 2008]. Organizations can therefore greatly benefit from insight into the maintainability of their spread-sheets, since it makes them aware of the latent risks and thus empowers them to take action to address them.

An assessment of the maintainability of a spreadsheet can be performed by an expert (or several) who would inspect the spreadsheet and give his opinion on how maintainable he thinks it is. Another option, however, is to estimate the maintainability based on *automated measurement*. The advantages of such an approach are clear:

1. **Objectivity:** A metric-based assessment is not influenced by the particular opinion of one or more experts, but based solely on facts: the measurement results;
2. **Repeatability:** Since the assessment is purely fact based, it can be more easily repeated or reproduced, as long as the measurements are deterministic and can be redone;
3. **Cost-effectiveness:** It is expected that an automated approach delivers a faster assessment than an expert. Also, an expert's time is more expensive than the time of a computer;
4. **Scalability:** Given the time and cost gain by using an automated approach, assessing a large set of spreadsheets is much less of a problem.

The Software Improvement Group (SIG) has created such an automated approach for evaluating the maintainability of software products (see [Heitlager, 2007] and [Baggen, 2010]). The TÜViT (a German certification party affiliated with the TÜV NORD Group) has accredited the SIG as an evaluation laboratory for the Trusted Product Maintainability certification scheme. The two key elements of the approach are a layered quality model that relates measurement results to high-level notions of quality, based on the ISO/IEC 9126 standard [ISO, 2001] and a continuously growing repository of evaluation results [Correia, 2008]. This *benchmark repository* allows us to study metrics' behavior in real-world software products and periodically recalibrate the quality model. Periodic calibration ensures that the quality model remains a reflection of the state of the art in software engineering.

We believe this methodology, which has shown its value both in practice as in some experimental studies (see [Luijten, 2010] and [Bijlsma, 2011]), can and should be applied to the evaluation of spreadsheet quality, in terms of *maintainability*.

The remainder of the paper is structured as follows. Section 2 presents some background on how the SIG's methodology works for software products. Section 3 defines the requirements for automated assessment of spreadsheet maintainability. Section 4 briefly describes the EUSES spreadsheet corpus. Section 5 describes the challenges in developing a maintainability model and curating a spreadsheet corpus to support it. Finally, Section 6 concludes the paper with some final remarks.

This paper aims at creating awareness for the issue of spreadsheet maintainability and to clearly lay out the requirements for a methodology that could support its automated assessment.

## 2 BACKGROUND

As mentioned in the previous section, two elements form the backbone of our approach: a quality model based on (automated) measurement and a repository of real-world instances of the object of study.

The SIG Quality Model is essentially a layered aggregation method to transform low-level, quantitative measurements into high-level ratings. A description can be found in the original publication [Heitlager, 2007]. The model has undergone some recent modifications since then, which can be found summarized in subsequent publications (*e.g.*, [Luijten, 2010], [Nugroho, 2011]).

In general terms, the application of the model involves the following steps:

1. Static analysis is performed on the source code to collect measurement data about the software system. Metrics are collected for the low-level system elements such as lines, units (e.g., methods or functions), and modules (e.g., files or classes). In the case of spreadsheets, low-level system elements can be cells, formulas, rows, and sheets;

2. Metrics for building blocks are mapped onto ratings for properties at the level of the entire software system, such as *volume*, *duplication*, *unit complexity*, etc. These ratings take values in the interval between 0.5 and 5.5, which can be rounded to an entire number of stars between one and five. (This constitutes a unitless ordinal scale that facilitates communication and comparison of quality results at the level of entire software systems.) The mapping functions for some properties are straightforward translations of system-level metrics to ratings. The remaining mapping functions make use of risk profiles as intermediate device. Each profile is a partition of the volume of the source code in each of four risk categories: low, moderate, high, and very high risk;

3. The ratings calculated per property are then aggregated to ratings for each of the four sub-characteristics of *maintainability* as defined in the ISO/IEC 9126, which are then finally aggregated to a rating for the maintainability of the whole software system. Which properties are deemed to influence which sub-characteristics is represented in a table with dependencies (for more on this mapping see [Correia, 2009]).

The mapping from metrics to ratings, described in point 2, is configured by a set of thresholds. These define the risk categories, as well as the mapping from either system-level metrics or risk profiles to ratings. The specific values for these thresholds have been calibrated based on statistical study of a large, representative sample of software systems [Alves, 2010], the SIG *benchmark repository*. This is crucial to the methodology, since it ensures that the model embodies realistic standards of quality. In order to make sure the model remains a reflection of the state of the art in software development, the repository is continuously updated with evaluation results, and re-calibration is performed periodically, as illustrated in Figure 1.

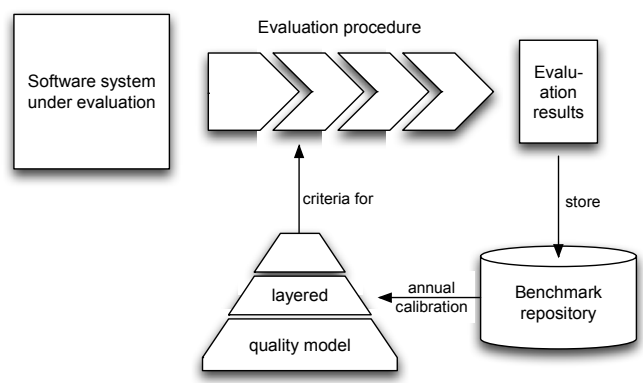

Figure 1 – Calibration cycle (adapted from [Baggen, 2010]).

## 4 WHAT IS NEEDED?

In order to adapt such an approach for spreadsheets, one can use a very similar set up, only tailoring it to account for the particular differences between these and traditional software systems. Nevertheless, this should be done carefully. We identify the following activities that need to be performed:

1. The goals of measurement and how we can achieve them through metrics need to be clearly defined. Many metrics for spreadsheets have already been defined in the literature (see [Bregar, 2004], [Hodnigg, 2008]), but in order to construct a model one needs to have present not just *how* to measure, but *what* (and *why*) to measure and *how to interpret* the results. We intend to apply the Goal Question Metric (GQM) approach [Basili, 1994] in order to achieve this. The GQM method provides a structured way of defining goals (in our case the ultimate goal is the assessment of spreadsheet maintainability), these goals are then broken down into different perspectives captured by questions and, finally, appropriate metrics are selected to provide data to answer the questions;

2. After a set of metrics is chosen, these should be investigated in a large, representative sample of spreadsheets. Namely, one needs to understand what the typical values of a given metric are, which extreme values can be found and whether they represent problematic situations or not. We intend to gather a large sample of spreadsheets, apply all the selected metrics consistently and perform statistical analysis, much in the fashion of [Alves, 2010].

The constitution of a representative repository of spreadsheets is very important for the implementation of the whole approach.

## 4 THE EUSES SPREADSHEET CORPUS

Since compiling a large and representative sample of spreadsheets can be quite challenging, we first decided to investigate what is available in the field of spreadsheet research. Apart from small sets used by different authors for their studies, we found only one large enough corpus, namely the EUSES Spreadsheet Corpus [Fisher, 2005].

This corpus is composed of 4498 spreadsheets and was set up by gathering spreadsheets from several different sources. Some curation was performed to exclude duplicated and unusable spreadsheets. The great majority (4401) was collected by querying the Google search engine[1]. The authors also calculated several metrics per spreadsheet, mostly counts of cell types for different categories of cells.

From this large corpus, we were interested only in spreadsheets that matched the following criteria:

- The spreadsheet should contain computations. Excluding spreadsheets without formula cells reduces the set to 1977 data points;
- The spreadsheet should contain referenced input cells. Excluding spreadsheet without those further reduces the set to 1609 data points.

---

[1] *http://www.google.com*

We performed some preliminary analysis on this subset and determined that at least 365 spreadsheets (approximately 23%) have more than 25 unique formulas (copy-equivalent), which is an indication that that there is a fare amount of non-trivial data points.

## 5 THE CHALLENGE

Even though the EUSES corpus is usable for a large-scale exploration of spreadsheet metrics, it has some disadvantages for its intended use in the context that we propose, namely:

- The corpus is most likely not representative of spreadsheets used in a professional context. The great majority comes from an internet search, where keywords such as "homework" or "grades" where used. The results of those queries are most likely spreadsheets used for personal management of information. Furthermore, most spreadsheets that implement some business value in an organization will not likely be publicly available on the internet;
- The corpus does not contain enough reliable meta-information to determine the actual representativeness of it, nor to allow for further analysis where grouping on different dimensions could provide some additional insight. It would be interesting, for example, to determine differences between spreadsheets used in different areas, developed by IT professionals versus end users, and so forth;
- There is no information on the legal permissions to analyze most spreadsheets in the corpus. The fact that they were publicly available through an internet search does not constitute legal permission of use;
- The age of the spreadsheets in the corpus is not clear. Furthermore, some 130 (out of the subset we considered) were found to be in a "BIFF5" format (format mainly used prior to Excel 97), which is an indication that they are quite old;
- The corpus is static: it was collected once and has not been updated since. We assume that spreadsheet development practices evolve and become better over time, this is one of the reasons why for software we re-calibrate the model periodically.

For all those reasons, we believe that there is a need for a spreadsheet repository compiled and managed in a similar manner to the SIG benchmark repository or, for example, the PROMISE[2] data set for machine learning research. We consider the following to be important characteristics of such a repository:

- **Curated:** The repository should contain spreadsheets that have been curated on arrival. This is important not only to verify the usability of the spreadsheet (much like it was done in the EUSES corpus), but also to determine their relevance and to make sure the proper meta-information is also included and appropriately handled;
- **Dynamic:** The repository should continuously grow, in order to reflect the state of the art in spreadsheet development. This means, of course, that it needs to be versioned to enable the proper replication of studies;
- **Open:** The repository should be easily available to the community and everybody should be allowed to contribute. If authors or maintainers contribute with their own spreadsheets, they can also provide extra meta-information to enrich the repository with. Some contributors might however have security concerns in disclosing their spreadsheets or other information, thus part of the repository could be restricted and subject to non-disclosure agreements for those who would like to use it;
- **Useful:** In order to motivate contribution from researchers, as well as practitioners, the repository must prove useful to them. For researchers, the use would be the availabil-ity of curated data for their studies. For any contributor, the repository could provide some general feedback on the quality of its contribution compared to the existing data. The whole repository could also be used to compile periodical, anonymous, "state of the art" reports. These are merely some examples.

---

[2] *http://promisedata.org/*

# 6 CONCLUSION

In this paper we have presented our view on how an automated assessment of spreadsheet maintainability could work, what are the requirements for it and what is still missing. Furthermore, we challenge the spreadsheet developer and user communities to contribute to a curated spreadsheet repository that can support a spreadsheet maintainability model (among potentially other approaches to spreadsheet quality). We are prepared to lead the effort in creating such a repository, but we find the involvement of the community fundamental for the success of such an endeavor. The initiative already has the collaboration of the SSaaPP project[3].

The approach we propose in this paper has already been successfully applied to assess the maintainability of software products. It is our conviction that with a carefully conducted study of spreadsheet metrics it is possible to select which are the most appropriate for building a similar maintainability model for spreadsheets.

We are open to other partnerships to support us in this effort, feedback on the whole project, but especially to contributions in terms of data points, so we leave the reader with a request: **may we have your spreadsheet?**

---

[3] *SpreadSheets as a Programming Paradigm (http://ssaapp.di.uminho.pt)*